# Arbitrary Measurement on Any Real-valued Probability Amplitude in Any Quantum System

Xu Guanlei*

(College of Computer and Information Engineering, Zhejiang Gongshang University, Hangzhou of China, 310018;

*xgl_86@163.com)

***Abstract*** — How to achieve an arbitrary real-valued probability amplitude in the general single-partite or multipartite quantum system without measuring any other quantum state's probability amplitude? How to achieve an arbitrary real-valued probability amplitude with the deterministic polynomial time's complexity under a small given error? In this paper, one novel quantum measurement scheme is proposed to solve these questions based on the idea of binary searching. First, the measurement algorithm with the exponential speed-up on the quantum state with one single qubit is well-designed. Then, the measurement algorithm is extended to the quantum states in the general multipartite quantum system and the special multipartite quantum system. The theoretical analysis proves that the proposed quantum measurement scheme has the performance in quantum information processing with twofold advantages: separable measurement and exponential speed up.

In quantum information processing based on quantum mechanics [1,2], the quantum measurement plays the key role for the information acquisition after quantum computation [3]. Up till now, various measurement manners such as general quantum measurement, POVM measurement, von Neumann measurement and weak measurement and so on have been widely exploited and used [3,4,5]. However, in order to achieve every probability amplitude in the quantum states, the traditional method is still the way of probability statistics [3]. For example, after a series of unitary operations we have the unknown quantum state $|\Psi\rangle$ with $|\Psi\rangle = \cos\alpha|0\rangle + \sin\alpha|1\rangle$ under the assumption of $0 \leq \cos\alpha, \sin\alpha \leq 1$ (or $0 \leq \alpha \leq \pi/2$) (there are many similar cases such as the representation of image quantum model FRQI [6], so this assumption holds true in practice in many cases), in order to know the value of $\alpha$, we have to prepare a large number (e.g., a large positive integer $N$) of quantum state $|\Psi\rangle$ for traditional measurement to obtain the probability amplitude $\sin\alpha$ by

$$p(|1\rangle) = \sin^2\alpha \approx \frac{N_1}{N}, \qquad (1)$$

where, $N$ is the total times of traditional quantum measurement, among which $N_1$ times' results are state $|0\rangle$ and $N_2$ times' results are state $|1\rangle$, $N = N_1 + N_2$, $p(|\varphi\rangle)$ denotes the probability of obtaining the quantum state $|\varphi\rangle$.

Therefore, we have the estimated angle $\alpha$ as follows

$$\alpha \approx \arcsin\sqrt{\frac{N_1}{N}}. \qquad (2)$$

If $N$ is very large and $N_1$ is small, then we will have $\alpha \approx \sqrt{\frac{N_1}{N}}$ in (2).

Moreover, with the increasing of $N$, the precision of $\alpha$ will increase as well. In principle, we hope $N \to +\infty$, then we will obtain the exact value of $\alpha$. However, in practice it is infeasible to execute $N \to +\infty$ times' quantum preparation and quantum measurement. Even so, we still need to execute a very large number of times' quantum preparation and traditional quantum measurement in actual quantum engineering.

In the same manner, in the multipartite quantum system for the quantum state $|\Psi\rangle = \sum_{i=0}^{2^n-1} \sin\alpha_i |i\rangle$ ($0 \leq \alpha_i \leq \pi/2$),

it is easily known that $p(|i\rangle) = \sin^2 \alpha_i \approx \frac{N_i}{N}$ and $N = \sum_{i=1}^{N} N_i$ with $\sum_{i=0}^{2^n-1} |i\rangle\langle i| = \mathbf{I}$ being $2^n \times 2^n$ matrix. In order to obtain the value of $\alpha_i$, we must execute a very large number of times' quantum preparation and traditional quantum measurement.

As shown in (2), we know the least measurement error (if the error occurs) during estimating $\alpha$ is about $\sqrt{\frac{1}{N_1}}$ (without loss of generality, we assume $N_1 = \frac{N}{2^n}$ in the average, where $N$ is the total traditional measurement times and $n$ is the total number of qubits).

Now if we require that our measurement error cannot be more than $\Delta e (> 0)$, then how many times' measurement are needed at least? According to the requirement, we have $\sqrt{\frac{1}{N_1}} = \sqrt{\frac{2^n}{N}} \leq \Delta e$. So we have

$$N \geq \frac{2^n}{(\Delta e)^2}. \qquad (3)$$

That is to say, in the average if we let the error of $\alpha$ be less than $\Delta e$, then we need to perform about $\frac{2^n}{(\Delta e)^2}$ times' traditional quantum measurement at least. At the same time, $\frac{2^n}{(\Delta e)^2}$ times' traditional quantum measurement means that we have to prepare the same quantum state $\frac{2^n}{(\Delta e)^2}$ times [3]. With the increasing of $n$, the traditional measurement and preparation complexity will increase in the speed of exponent.

Then is there one method that executes a much less number of times' (e.g., deterministic polynomial times) quantum preparation and quantum measurement to obtain the value of $\alpha$ with high precision?

In the follows, we will give the answer: Yes, the novel quantum measurement scheme will be given in this paper. At first, we will yield the proposed algorithm for the single-partite quantum state. Then we will show the extended algorithm for the case of the multipartite quantum states so that we can achieve the measurement of the probability amplitude for any quantum state in the multipartite quantum system with much less complexity. At the last, we will conclude our work and show the future work.

## 1. For the case of single-partite quantum system

In quantum information processing, the unitary operators such as NOT gate $\mathbf{X}$, Hadamard gate $\mathbf{H}$ and $\mathbf{Z}$ gate and so on (shown in (4)) are widely used in various quantum computation [3].

$$\mathbf{H} = \frac{1}{\sqrt{2}} \begin{bmatrix} 1 & 1 \\ 1 & -1 \end{bmatrix}, \mathbf{Z} = \sigma_z = \begin{bmatrix} 1 & 0 \\ 0 & -1 \end{bmatrix}, \mathbf{X} = \sigma_x = \begin{bmatrix} 0 & 1 \\ 1 & 0 \end{bmatrix}. \qquad (4)$$

As the generalized extension of the three unitary operators in (4), we have the generalized unitary operator shown as follows:

$$\mathbf{C}(2\beta) = \begin{bmatrix} \cos\beta & \sin\beta \\ \sin\beta & -\cos\beta \end{bmatrix}. \qquad (5)$$

Now perform the unitary rotation operator $\mathbf{C}(2\beta)$ on the original state $|0\rangle$, we have

$$|\Phi\rangle_0 = \mathbf{C}(2\beta)|0\rangle = \cos\beta|0\rangle + \sin\beta|1\rangle = \begin{bmatrix} \cos\beta \\ \sin\beta \end{bmatrix}. \qquad (6)$$

Let $|\Phi\rangle_1 = |\Psi\rangle \otimes |\Phi\rangle_0 = \begin{bmatrix} \cos\alpha \\ \sin\alpha \end{bmatrix} \otimes \begin{bmatrix} \cos\beta \\ \sin\beta \end{bmatrix} = \begin{bmatrix} \cos\alpha\cos\beta \\ \cos\alpha\sin\beta \\ \sin\alpha\cos\beta \\ \sin\alpha\sin\beta \end{bmatrix}$ and use the unitary operator $\mathbf{A}$ on $|\Phi\rangle_1$, we have

$$|\Phi\rangle_2 = \mathbf{A}|\Phi\rangle_1 = \mathbf{A} \cdot \begin{bmatrix} \cos\alpha\cos\beta \\ \cos\alpha\sin\beta \\ \sin\alpha\cos\beta \\ \sin\alpha\sin\beta \end{bmatrix} = \frac{1}{\sqrt{2}} \begin{bmatrix} \cos\alpha\cos\beta - \sin\alpha\sin\beta \\ \cos\alpha\sin\beta + \sin\alpha\cos\beta \\ \sin\alpha\cos\beta - \cos\alpha\sin\beta \\ \sin\alpha\sin\beta + \cos\alpha\cos\beta \end{bmatrix} = \frac{1}{\sqrt{2}} \begin{bmatrix} \cos(\alpha+\beta) \\ \sin(\alpha+\beta) \\ \sin(\alpha-\beta) \\ \cos(\alpha-\beta) \end{bmatrix}, \quad (7)$$

where $\mathbf{A} = \frac{1}{\sqrt{2}} \begin{bmatrix} 1 & 0 & 0 & -1 \\ 0 & 1 & 1 & 0 \\ 0 & 1 & -1 & 0 \\ 1 & 0 & 0 & 1 \end{bmatrix}$, $0 \leq \alpha, \beta \leq \pi/2$.

Via equation (7), we have the following quantum state

$$|\Phi\rangle_2 = \frac{1}{\sqrt{2}} \left[ \cos(\alpha+\beta)|00\rangle + \sin(\alpha+\beta)|01\rangle + \sin(\alpha-\beta)|10\rangle + \cos(\alpha-\beta)|11\rangle \right]. \quad (8)$$

For the quantum state $|\Phi\rangle_2$, now we set the projection operator $\mathbf{P} = |10\rangle\langle 10|$, and perform von Neumann measurement, we will have the quantum state after measurement

$$|\Phi\rangle_3 = \frac{\mathbf{P}|\Phi\rangle_2}{\sqrt{\langle\Phi|_2 \mathbf{P}|\Phi\rangle_2}} = \begin{cases} 0, & \alpha = \beta \\ +|10\rangle, & \alpha > \beta \\ -|10\rangle, & \alpha < \beta \end{cases}. \quad (9)$$

Clearly, from the measured quantum state in (9), we can easily know the relation between $\alpha$ and $\beta$.

If $|\Phi\rangle_3 = 0$, then we directly obtain the value of $\alpha$, i.e., $\alpha = \beta$. If $|\Phi\rangle_3 = +|10\rangle$, we know $\alpha > \beta$. If $|\Phi\rangle_3 = -|10\rangle$, we know $\alpha < \beta$.

Thus, through the relation between $\alpha$ and $\beta$, we can change the value of $\beta$ and repeat the operations shown in (6)-(9) $m$ times, we can obtain the result by

$$\alpha = \beta \pm \Delta e \text{ with the error } \Delta e \leq \frac{\pi}{2^{m+1}}. \quad (10)$$

Here, we note that after $m$ times' operations the max error is $\frac{\pi/2}{2^m} = \frac{\pi}{2^{m+1}}$, so we have $\Delta e \leq \frac{\pi}{2^{m+1}}$. Therefore, after $m$ times' quantum operations, we obtain $\alpha \approx \beta$ with the max error $\frac{\pi}{2^{m+1}}$. With the increasing of $m$, the error will attenuate in the exponential speed, e.g., $m = 10 \rightarrow \Delta e \leq \frac{\pi}{2^{m+1}} \approx 1.5 \times 10^{-3}$, $m = 20 \rightarrow \Delta e \leq \frac{\pi}{2^{m+1}} \approx 10^{-6}$.

Thus, we give our proposed algorithm (Algorithm I) shown as follows.

**The input**: the unknown state $|\Psi\rangle = \cos\alpha|0\rangle + \sin\alpha|1\rangle$, the original state $|0\rangle$, the unitary operators $\mathbf{C}(2\beta)$ and $\mathbf{A}$, the projection operator $\mathbf{P}$, $\beta_1 = 0$, $\beta_2 = \pi/2$, the angle $\beta$ selected randomly in $[0, \pi/2]$, $0 \leq \alpha, \beta \leq \pi/2$, the required estimation error $\Delta e$ of $\alpha$.

**The output**: the estimated value of $\alpha$ with error $\Delta e \leq \frac{\pi}{2^{m+1}}$.

**The complexity**: $O(m) \leq O\left(1 + \log_2 \frac{\pi}{\Delta e}\right) \approx O\left(\log_2 \frac{\pi}{\Delta e}\right)$.

**The procedure**:

**Step 1**: Prepare $|0\rangle$ and $|\Psi\rangle = \cos\alpha|0\rangle + \sin\alpha|1\rangle$.

**Step 2:** $\mathbf{C}(2\beta)|0\rangle \to |\Phi\rangle_0 = \begin{bmatrix} \cos\beta \\ \sin\beta \end{bmatrix}$.

**Step 3:** $|\Psi\rangle \otimes |\Phi\rangle_0 \to |\Phi\rangle_1 = \begin{bmatrix} \cos\alpha\cos\beta \\ \cos\alpha\sin\beta \\ \sin\alpha\cos\beta \\ \sin\alpha\sin\beta \end{bmatrix}$.

**Step 3:** $\mathbf{A}|\Phi\rangle_1 \to |\Phi\rangle_2 = \dfrac{1}{\sqrt{2}} \begin{bmatrix} \cos(\alpha+\beta) \\ \sin(\alpha+\beta) \\ \sin(\alpha-\beta) \\ \cos(\alpha-\beta) \end{bmatrix}$.

**Step 4:** $\dfrac{\mathbf{P}|\Phi\rangle_2}{\sqrt{\langle\Phi|_2 \mathbf{P}|\Phi\rangle_2}} \to |\Phi\rangle_3 = \begin{cases} 0 \Rightarrow \alpha = \beta \\ +|10\rangle \Rightarrow \alpha > \beta \\ -|10\rangle \Rightarrow \alpha < \beta \end{cases}$.

**Step 5:** If $|\Phi\rangle_3 = 0$, then $\beta \to \alpha$ and stop; Otherwise, let $\begin{cases} \beta_1 = \beta, & \text{if } \alpha > \beta \\ \beta_2 = \beta, & \text{if } \alpha < \beta \end{cases}$ and $\beta = \begin{cases} \dfrac{\beta+\beta_2}{2}, & \text{if } \alpha > \beta \\ \dfrac{\beta+\beta_1}{2}, & \text{if } \alpha < \beta \end{cases}$, then

**repeat steps (2)-(5)** $m$ **times and** $\beta \to \alpha$, **stop.**

It has been shown that this proposed algorithm makes full use of the idea of binary searching to obtain the approximation of $\alpha$ so that only about $m$ times' operations are needed. Indeed, from this algorithm we also know that there are four steps in every repetition period, so the actual operations are about $4m$ times.

## 2. For the case of multipartite quantum system

### 2.1 General case

In the case of single-partite quantum state, there is only one single qubit. However, in the multipartite quantum system, we have to process the multipartite quantum states with more qubits.

For any given multipartite quantum states as follows

$$|\Psi\rangle = \sum_{i=0}^{2^n-1} a_i |i\rangle, \qquad (10)$$

where, $\sum_{i=0}^{2^n-1} |a_i|^2 = 1$, meanwhile we assume $0 \le a_i \le 1$ (as is often true, e.g., in the FRQI quantum image representation model [6], all the probability amplitudes are non-negative real-valued ones).

At the same time, for the any given quantum state $|k\rangle$ ( $0 \le k \le 2^n - 1$ ), we can rewrite (10) as follows

$$|\Psi\rangle = \cos\theta|\phi\rangle + \sin\theta|k\rangle, \qquad (11)$$

where, $\cos\theta|\phi\rangle \triangleq \sum_{i=0, i\ne k}^{2^n-1} a_i |i\rangle$, $\sin\theta = a_k$, $|k\rangle\langle k| + |\phi\rangle\langle\phi| = \mathbf{I}$, $0 \le \theta \le \pi/2$.

Similarly, the rest of our second algorithm has the similar operations as Algorithm I as follows:

$$|\Phi\rangle_0 = \mathbf{C}(2\beta)|0\rangle = \cos\beta|0\rangle + \sin\beta|1\rangle = \begin{bmatrix} \cos\beta \\ \sin\beta \end{bmatrix}. \qquad (12)$$

$$|\Phi\rangle_1 = \begin{bmatrix} \cos\theta \\ \sin\theta \end{bmatrix} \otimes \begin{bmatrix} \cos\beta \\ \sin\beta \end{bmatrix} = \begin{bmatrix} \cos\theta\cos\beta \\ \cos\theta\sin\beta \\ \sin\theta\cos\beta \\ \sin\theta\sin\beta \end{bmatrix}. \qquad (13)$$

$$|\Phi\rangle_2 = \mathbf{A}|\Phi\rangle_1 = \frac{1}{\sqrt{2}} \begin{bmatrix} \cos(\theta+\beta) \\ \sin(\theta+\beta) \\ \sin(\theta-\beta) \\ \cos(\theta-\beta) \end{bmatrix}, \quad \mathbf{A} = \frac{1}{\sqrt{2}} \begin{bmatrix} 1 & 0 & 0 & -1 \\ 0 & 1 & 1 & 0 \\ 0 & 1 & -1 & 0 \\ 1 & 0 & 0 & 1 \end{bmatrix} \text{ and } 0 \le \theta, \beta \le \pi/2. \quad (14)$$

That is, we finally obtain the following quantum state for the multipartite quantum system

$$|\Phi\rangle_2 = \frac{1}{\sqrt{2}}\left[\cos(\theta+\beta)|\phi\rangle|0\rangle + \sin(\theta+\beta)|\phi\rangle|1\rangle + \sin(\theta-\beta)|k\rangle|0\rangle + \cos(\theta-\beta)|k\rangle|1\rangle\right]. \quad (15)$$

For the quantum state $|\Phi\rangle_2$, we set new projection operator $\mathbf{P} = |k\rangle|0\rangle\langle k|\langle 0|$ (which is different from that shown in Algorithm I) and perform von Neumann measurement on $|\Phi\rangle_2$ to obtain the following quantum state

$$|\Phi\rangle_3 = \frac{\mathbf{P}|\Phi\rangle_2}{\sqrt{\langle\Phi|_2 \mathbf{P}|\Phi\rangle_2}} = \begin{cases} 0, & \theta = \beta \\ +|k\rangle|0\rangle, & \theta > \beta \\ -|k\rangle|0\rangle, & \theta < \beta \end{cases}. \quad (16)$$

In the same way, the proposed algorithm (Algorithm II) is achieved as follows.

**The input**: the unknown state $|\Psi\rangle = \sum_{i=0}^{2^n-1} a_i |i\rangle$, the original state $|0\rangle$ and the given quantum state $|k\rangle$, the unitary operators $\mathbf{C}(2\beta)$, the unitary operator $\mathbf{A}$ and the new projection operator $\mathbf{P}$, $\beta_1 = 0$, $\beta_2 = \pi/2$, the angle $\beta$ given randomly in $[0, \pi/2]$, $0 \le \theta, \beta \le \pi/2$, the given max estimation error $\Delta e$ of $\theta$.

**The output**: the estimated value of probability amplitude of $|k\rangle$ (turn to $\theta$ with max error $\frac{\pi}{2^{m+1}}$).

**The complexity**: $O(m) \approx O\left(\log_2 \frac{\pi}{|\Delta e|}\right)$.

**The procedure**:

Step 1: Prepare $|0\rangle$ and $|\Psi\rangle = \cos\theta|\phi\rangle + \sin\theta|k\rangle$.

Step 2: $\mathbf{C}(2\beta)|0\rangle \rightarrow |\Phi\rangle_0 = \begin{bmatrix} \cos\beta \\ \sin\beta \end{bmatrix}$.

Step 3: $|\Psi\rangle \otimes |\Phi\rangle_0 \rightarrow |\Phi\rangle_1 = \begin{bmatrix} \cos\theta\cos\beta \\ \cos\theta\sin\beta \\ \sin\theta\cos\beta \\ \sin\theta\sin\beta \end{bmatrix}$.

Step 3: $\mathbf{A}|\Phi\rangle_1 \rightarrow |\Phi\rangle_2 = \frac{1}{\sqrt{2}} \begin{bmatrix} \cos(\theta+\beta) \\ \sin(\theta+\beta) \\ \sin(\theta-\beta) \\ \cos(\theta-\beta) \end{bmatrix}$.

Step 4: $\frac{\mathbf{P}|\Phi\rangle_2}{\sqrt{\langle\Phi|_2 \mathbf{P}|\Phi\rangle_2}} \rightarrow |\Phi\rangle_3 = \begin{cases} 0 \Rightarrow \theta = \beta \\ +|k\rangle|0\rangle \Rightarrow \theta > \beta \\ -|k\rangle|0\rangle \Rightarrow \theta < \beta \end{cases}$.

Step 5: If $|\Phi\rangle_3 = 0$, then $\beta \rightarrow \theta$ and stop; Otherwise, let $\begin{cases} \beta_1 = \beta, & \text{if } \theta > \beta \\ \beta_2 = \beta, & \text{if } \theta < \beta \end{cases}$ and $\beta = \begin{cases} \dfrac{\beta+\beta_2}{2}, & \text{if } \theta > \beta \\ \dfrac{\beta+\beta_1}{2}, & \text{if } \theta < \beta \end{cases}$, then repeat steps (2)-(5) $m$ times and $\beta \rightarrow \theta$, stop.

The max difference between Algorithm II and Algorithm I lies in twofold: one is that the new projection operator $\mathbf{P}=|k\rangle|0\rangle\langle k|\langle 0|$ in Algorithm II is very different from that of Algorithm I; the another one is the unknown state $|\Psi\rangle=\cos\theta|\phi\rangle+\sin\theta|k\rangle$ which has more qubits along with more quantum states.

Algorithm II tells us that we can achieve any probability amplitude in all the $2^n$ quantum states through about $O\left(\log_2\frac{\pi}{\Delta e}\right)$ times' quantum operations for the given state $|k\rangle$. In addition, the complexity $O\left(\log_2\frac{\pi}{\Delta e}\right)$ is free of the total number of quantum states: $2^n$.

Moreover, on one hand, the traditional measurement methods cannot only measure one certain single probability amplitude in all the $2^n$ quantum states, instead, they have to measure all the states simultaneously to obtain all the probability amplitudes.

On the other hand, if we want to know all the probability amplitudes in the $2^2$ quantum states, we need about $O\left(2^n\cdot\log_2\frac{\pi}{\Delta e}\right)$ times' quantum operations (the same times' quantum state preparation are needed at the same time), which will be much less than the traditional measurement methods with the complexity $O\left(\frac{2^{2n}}{(\Delta e)^2}\right)$ at least (the same times' quantum state preparation are needed as well).

If we set $\log_2\frac{\pi}{\Delta e}\approx m$, then we need about $O(m\cdot 2^n)$ times' quantum operations corresponding to the complexity $O(2^{2m}\cdot 2^{2n})$ of the traditional quantum measurement.

## 2.2 Special case

For the quantum state shown in (10), if we are able to rewrite it as follows

$$|\Psi\rangle=\bigotimes_{i=1}^{n}\left(\cos\theta_i|0\rangle+\sin\theta_i|1\rangle\right), \quad (17)$$

where, $0\leq\theta_i\leq\pi/2$, then we will further speed up the quantum measurement for the multipartite quantum system.

Based on the tensor product form in (17), we have

$$|\Psi\rangle=\cos\theta_l|0\rangle|\boldsymbol{\omega}\rangle+\sin\theta_l|1\rangle|\boldsymbol{\omega}\rangle, \quad (18)$$

where, $|\boldsymbol{\omega}\rangle=\bigotimes_{i=1,i\neq l}^{n}\left(\cos\theta_i|0\rangle+\sin\theta_i|1\rangle\right)$ and $1\leq l\leq n$.

Therefore, set $\mathbf{P}=|\boldsymbol{\omega}\rangle|1\rangle\langle\boldsymbol{\omega}|\langle 1|$, we only need perform $n$ times' Algorithm II in (18), we can obtain all the $\theta_i$, thus our total complexity is about $O(m\cdot n)$ corresponding to the complexity $O(2^{2m}\cdot 2^n)$ of the traditional quantum measurement.

## 3. Conclusion and future work

This paper proposed a novel quantum measurement scheme instead of the traditional quantum measurement for the real-valued probability amplitudes. The proposed algorithm makes full use of the idea of binary searching to approach the true value in the real-valued probability amplitude. The main contributions of our work are twofold: one is that the arbitrary real-valued probability amplitude in any quantum system can be achieved free of measuring any other quantum state's probability in our scheme but the traditional counterpart fails to do so; another one is that our scheme's complexity is about $O(m)$ ($\log_2(\pi/\Delta e)\triangleq m$) according to the complexity $O(2^{2m+n})$ of the traditional measurement under the same error for the measurement of one real-valued probability amplitude, in addition we need about $O(m\cdot 2^n)$ ($O(m\cdot n)$ in the special case) times' quantum operations corresponding to the complexity $O(2^{2m}\cdot 2^{2n})$ of the traditional quantum measurement for the measurement of all

$2^n$ states. To our best knowledge, our proposed measurement scheme in this paper is completely novel and there has been no any report covering such measurement idea.

However, our proposed scheme fails to measure the complex-valued probability amplitudes in the quantum states because our method is confined to the real-valued ones [6][7]. In other words, if $|\Psi\rangle = \cos\alpha|0\rangle + e^{j\zeta}\sin\alpha|1\rangle$ (where, $j$ is the complex unit and $\zeta$ is real-valued phase), then we can only obtain the term of $\sin\alpha$ without $e^{j\zeta}$. Therefore, in future, we will extend our idea to the work on how to obtain the complex-valued ones. At the same time, we don't take into account of the errors resulted by the various unitary transforms during the quantum state preparation and quantum computation in this paper, so the future work will also include the influence analysis on these errors.

**Acknowledgement:** The work in this paper is fully supported by the NSFCs (6207012505, 61771020, 61471412) and 2019KD0AC02.

## REFERENCES

[1] Merzbacher E . Quantum Mechanics[J]. Physics Today, 1998, 24(6):49-50.

[2] Lawden D F , Sposito G . The Mathematical Principles of Quantum Mechanics[J]. Physics Today, 1969, 22(5):82-85.

[3] Nielsen M A , Chuang I L . Quantum computation and quantum information[J]. Mathematical Structures in Computer Science, 2007, 17(6):1115-1115.

[4] Steinberg, Aephraim M . Quantum measurement: A light touch[J]. Nature, 2010, 463(7283):890-891.

[5] Palacios-Laloy A , Mallet F , Nguyen F , et al. Experimental violation of a Bell's inequality in time with weak measurement[J]. Nature physics, 2010, 6(6):442-447.

[6] Le P Q , Dong F , Hirota K . A flexible representation of quantum images for polynomial preparation, image compression, and processing operations[J]. Quantum information processing, 2011, 10(1):63-84.

[7] Xu G , Xu X , Wang X , et al. Order-encoded quantum image model and parallel histogram specification[J]. Quantum Information Processing, 2019, 18(11), DOI: 10.1007/s11128-019-2463-7.